\documentclass[journal = jpcc, manuscript=article,layout = onecolumn]{achemso}

\setkeys{acs}{articletitle = true}
\usepackage{color,soul}
\usepackage{amsmath,amssymb}
\usepackage{subcaption}
\usepackage{xcolor}
\usepackage{graphicx}
\usepackage{dcolumn}
\usepackage{bm}
\usepackage{tikz}

\definecolor{MyBlue}{RGB}{55,126,184}

\title{Salt-in-Ionic-Liquid Electrolytes: Ion Network Formation and Negative Effective Charges of Alkali Metal Cations}

\author{Michael McEldrew}
\affiliation{Department of Chemical Engineering, Massachusetts Institute of Technology, Cambridge, MA, USA}

\author{Zachary A. H. Goodwin}
\affiliation{Department of Chemistry, Imperial College of London, Molecular Sciences Research Hub, White City Campus, Wood Lane, London W12 0BZ, UK}
\alsoaffiliation{Thomas Young Centre for Theory and Simulation of Materials, Imperial College of London, South Kensington Campus, London SW7 2AZ, UK}

\author{Nicola Molinari}
\affiliation{John A. Paulson School of Engineering and Applied Sciences, Harvard University, Cambridge, Massachusetts 02138, United States}

\author{Boris Kozinsky}
\affiliation{John A. Paulson School of Engineering and Applied Sciences, Harvard University, Cambridge, Massachusetts 02138, United States}

\author{Alexei A. Kornyshev}
\email{a.kornyshev@imperial.ac.uk}
\affiliation{Department of Chemistry, Imperial College of London, Molecular Sciences Research Hub, White City Campus, Wood Lane, London W12 0BZ, UK}
\alsoaffiliation{Thomas Young Centre for Theory and Simulation of Materials, Imperial College of London, South Kensington Campus, London SW7 2AZ, UK}
\alsoaffiliation{Institute of Molecular Science and Engineering, Imperial College of London, South Kensington Campus, London SW7 2AZ, UK}

\author{Martin Z. Bazant}
\email{bazant@mit.edu}
\affiliation{Department of Chemical Engineering, Massachusetts Institute of Technology, Cambridge, MA, USA}
\alsoaffiliation{Department of Mathematics, Massachusetts Institute of Technology, Cambridge, MA, USA}

\date{\today}
             
\begin{document}
\newpage

\begin{abstract}
Salt-in-ionic liquid electrolytes have attracted significant attention as potential electrolytes for next generation batteries largely due to their safety enhancements over typical organic electrolytes.  However, recent experimental and computational studies have shown that under certain conditions alkali cations can migrate in electric fields as if they carried a net negative effective charge. In particular, alkali cations were observed to have negative transference numbers at small mole fractions of alkali metal salt that revert to the expected net positive transference numbers at large mole fractions. Simulations have provided some insights into these observations, where the formation of asymmetric ionic clusters, as well as a percolating ion network could largely explain the anomalous transport of alkali cations. However, a thermodynamic theory that captures such phenomena has not been developed, as ionic associations were typically treated via the formation of ion pairs. The theory presented herein, based on the classical polymer theories, describes thermoreversible associations between alkali cations and anions, where the formation of large, asymmetric ionic clusters and a percolating ionic network are a natural result of the theory.  Furthermore, we present several general methods to calculate the effective charge of alkali cations in ionic liquids. We note that the negative effective charge is a robust prediction with respect to the parameters of the theory, and that the formation of a percolating ionic network leads to the restoration of net positive charges of the cations at large mole fractions of alkali metal salt. Overall, we find excellent qualitative agreement between our theory and molecular simulations in terms of ionic cluster statistics and the effective charges of the alkali cations.
\end{abstract}
   
\maketitle

\newpage

\section{Introduction}
Ionic liquids (ILs) are experiencing a surge in scientific and technological interest in a variety of fields owing to their tailorable nature and unique physicochemical properties~\cite{Welton1999,Hermann2008,Welton2011}.
In the energy community, the low volatility, low flammability, and high chemical and thermal stabilities of IL electrolytes~\cite{Fedorov2014,dusastre2010materials,macfarlane2014energy,watanabe2017application} make them attractive alternatives to conventional organic solvents in applications ranging from electric double-layer transistors to batteries~\cite{dokko2013solvate,lewandowski2009ionic}, solar cells~\cite{zakeeruddin2009solvent}, supercapacitors~\cite{balducci2004ionic, balducci2007high}, advanced fuel cells~\cite{noda2003bronsted}, and advanced carbon capture~\cite{ramdin2012state}.
Specifically, for battery applications, electrolytes solutions are typically composed of a IL doped with an alkali metal, most commonly lithium, salt~\cite{dokko2013solvate,lewandowski2009ionic}.
As the charge and discharge rate capabilities of a battery are largely determined by its transport properties, namely ionic conductivity and alkali-cation transference number, the design of next-generation batteries relies on the development of fast ion-conducting electrolytes~\cite{schmuch2018performance, xu2014electrolytes, balduccicritical}.
%

%
%
%

%

\indent Intuitively, due to the proportionality between number of charge carriers and ionic conductivity, a route to optimise transport properties of alkali metal cations in ILs electrolytes is to use high concentrations of alkali metal salt.
This natural strategy, however, results in significant ionic correlations~\cite{chen2016elucidation,qiao2018supramolecular, molinari2019transport,Kubisiak2020, lourenco2021theoretical} stemming from the strong Coulombic interactions of both the IL solvent and the alkali-ion salt.
Recent experimental works adopting electrophoretic NMR to measure ionic mobilities~\cite{gouverneur2015direct,gouverneur2018negative,Marc2021}, as well as computational works applying concentrated solution theory to molecular dynamics (MD) simulations~\cite{molinari2019transport, molinari2019general,chen2016elucidation,Kubisiak2020} unveiled the surprising transport anomaly of a negative alkali-ion transference number arising in this strongly correlated system.
To explain such an observation, the existence of ionic agglomerates containing more IL-anions than alkali cations was postulated~\cite{gouverneur2018negative}, and the molecular resolution of atomistic simulations confirmed this hypothesis~\cite{molinari2019general}.
While experiments are limited by the high viscosity of highly-concentrated IL electrolytes, MD simulations extended the analysis to high alkali-salt molar fractions, showing the ubiquitous tendency of these systems to percolate into fully-connected networks, and of the alkali-ion transference number to reverse sign and approach $+1$~\cite{molinari2019general,molinari2020chelation,Ackermann2021}.
Both communities took an additional step towards rationalising these puzzling observations by independently measuring the alkali-cation effective charge, i.e., the time averaged charge carried by alkali-cation-containing clusters~\cite{gouverneur2018negative, molinari2019general}.
It appeared that the alkali-cation effective charge, $q^{eff}$, qualitatively follows the behaviour of the alkali-cation transference number\cite{molinari2019general}.
The $q^{eff}$ is found to be negative for low-to-moderate alkali salt molar fractions, concomitant with the existence of small and asymmetrical IL-anion/alkali-cation clusters.
For higher alkali salt molar fractions, $q^{eff}$ abruptly reverses sign to positive values, while the cluster population percolates to a single network.
The latter is immobile, and only rare, non-clustered Li-ions appear to dominate the charge transport. 

While this experimental and computational understanding is helpful to guide the formulation of electrolyte design rules~\cite{molinari2020chelation,Ackermann2021}, it lacks the support of a thermodynamic theory of ion clustering that can provide a conceptual, mathematical framework for the phenomena observed in experiments and simulations. 
Some preliminary thermodynamic theories have attempted to capture ionic associations in super-concentrated electrolytes through ion pair formation~\cite{lee2014room} or by neglecting correlations and only accounting for free ions~\cite{Goodwin2017}.
While these approaches can be useful for modelling certain properties of super-concentrated electrolytes~\cite{Yufan2020}, such as transport~\cite{feng2019free} and differential capacitance~\cite{Chen2017} properties of neat ILs, their assumptions prevent the description of salt-in-IL systems where clusters of more than two ions gives rise to net negative effective charges of the alkali cation.
It is the explicit description of larger clusters that is required to describe salt-in-IL. 

\indent In this article, we aim to fill this gap by developing a thermodynamically consistent theory for ionic clustering and network formation of salt-in-ionic liquids. 
This theory is based on the classical theories of thermoreversible association and gelation in polymer mixtures, which some of us have recently modified and extended for applications in super-concentrated electrolytes \cite{mceldrew2020theory,mceldrew2020corr,mceldrew2021ion}. 
Our theory naturally depends on only a handful of parameters, and importantly it is extended here to depend on the mole fraction of alkali metal salt that is added to an IL, which shares the same anion as the alkali metal salt. 
Using this theory, we are able to compute the distribution of clusters, discern the onset of a percolating ionic network, and calculate the effective charge of clustered ions.
In order to test the theory, we use molecular simulations of IL/alkali metal salt mixtures, EmimBF$_4$/LiBF$_4$ and EmimPF$_6$/LiPF$_6$, to obtain its few key parameters and check the consistency of its assumptions.
Using these parameters, we compare the cluster distributions, gelation point, and the effective charge predicted from our theory against those computed from MD, and find remarkably good agreement between theory and simulations. The agreement between theory and simulations of these electrolytes is a benchmark to warrant its application to novel systems. Our theory gives a framework to understand and provide intuition for complex ionic liquid-based electrolytes. As these mixtures become more and more complex with the incorporation of solvent molecules~\cite{chen202063}, chelating agents~\cite{Ackermann2021,molinari2020chelation}, or co-anions~\cite{Suo2016,wang2020highly}, theoretical guidance can be invaluable in sweeping through volumes of design space that even high-throughput experimentation and molecular simulation cannot cover.



%

\section{Theory}

In modelling ionic clusters of alkali metal salts in ILs, the theoretical approach outlined in Refs.~\citenum{mceldrew2020theory,mceldrew2020corr,mceldrew2021ion} is followed. We treat the mixture as an incompressible lattice fluid containing a polydisperse mixture of alkali cation--anion Cayley (loop-less) tree clusters, as well as IL cations which are assumed not to participate in ionic associations. The IL cation is denoted with subscript, $c_{1}$, the alkali cation with subscript, $c_{2}$, and the anion, shared between the IL and alkali metal salt, with subscript, $a$. An alkali cation can associate to at most $f_{c_{2}}$ anions, and anions can associate to at most $f_a$ alkali cations, which are referred to as the functionalities of the respective ions.

The assumption of Cayley tree clusters for super-concentrated electrolytes is not always a good approximation. For instance, electrolytes containing more ``kosmotropic" salts, such as NaCl, tend to form ordered clusters with substantial numbers of loops. Such crystalline clusters cannot be well described with the presented theory, and would instead require a theory to describe precipitation of the solid phase. However, electrolytes containing ``chaotropic" salts, such as LiTFSI, form highly branched and disordered clusters with practically no intra-cluster loops~\cite{mceldrew2021ion}. Therefore, the assumption of Cayley tree clusters is reasonable for many IL-based systems, especially for ones with more bulky, asymmetric and chaotropic anions.

Our model also allows for the formation of a percolating ion network. In order to do this, the model partitions the electrolyte into a ``sol" and ``gel" as is analogously defined in the works of Tanaka \textit{et al.}~\cite{tanaka1989,tanaka1990thermodynamic,tanaka1994,tanaka1995,ishida1997,tanaka1998,tanaka1999,tanaka2002} in modelling  thermoreversible association and gelation in polymer mixtures. Here, the gel is defined as the part of the electrolyte that is incorporated in the percolating ion network, and the sol is defined as the part of the electrolyte that is not incorporated in the percolating ion network. In this way, the gel and percolating ion network can be used interchangeably. Here, however, to avoid any preconception, we only refer to the percolating ion network as a ``gel" in our mathematical notation (keeping in line with the notation of Tanaka's many works).

As previously mentioned we neglect explicit associations involving IL cations, and this is motivated by the expectation that the interactions between IL cations and anions will be \emph{much weaker} than the interactions between alkali cations and anions. Indeed, in Ref.~\citenum{mceldrew2020corr}, the association constant between anions and cations in all of the studied ILs were found to be less than one. Whereas, as shown in Ref.~\citenum{mceldrew2021ion}, the associations between lithium cations and IL anions were found to be much larger than one, and thus more important to explicitly model. However, in order to avoid entirely neglecting interactions between IL cations and anions, we model the IL cations as interacting with the open association sites of the anions via regular solution interactions~\cite{goodwin2017mean,Goodwin2017}. These regular solution interactions provide a simple mean-field description of the competition between alkali metal and IL cations for interactions with the anions, without incurring the mathematical complexity of modelling three-component ionic clusters--the combinatorics of which (as far as the authors are aware) has yet to be resolved mathematically.

A Flory-like lattice fluid free energy of mixing~\cite{flory1942thermodynamics}, used extensively for polydisperse mixtures of thermoreversibly-associating polymer mixtures~\cite{tanaka1989,tanaka1990thermodynamic,tanaka1994,tanaka1995,ishida1997,tanaka1998,tanaka1999,tanaka2002}, is employed
\begin{align}
\beta \Delta F &= N_{c_{1}}\ln \left(\phi_{c_{1}}\right) + \sum_{lm} \left[N_{lm}\ln \left( \phi_{lm} \right)+N_{lm}\Delta_{lm}\right] \nonumber\\
&+ \beta \chi \phi_{c_1}\sum_{lm}\left[(f_{a}m-m-l+1)N_{lm} \right] \nonumber \\
&+ \Delta^{gel}_{c_{2}} N^{gel}_{c_{2}} + \Delta^{gel}_{a} N^{gel}_a. 
\label{eq:F}
\end{align}
Here $\beta = 1/k_BT$ is inverse thermal energy; $N_{c_{1}}$ and $\phi_{c_{1}}$ are the mole number and volume fraction, respectively, of the IL cation; $N_{lm}$ and $\phi_{lm}$ are the number and volume fraction, respectively, of rank $lm$ clusters with $l$ alkali cations and $m$ anions; $\Delta_{lm}$ is the free energy of formation of a rank $lm$ cluster, which can have contributions from the combinatorial entropy, bonding energy, and configurational entropy; $\chi$ is the dimensionless regular solution interaction parameter capturing the mean-field enthalpy of mixing between IL cations and the open association sites on anions; $\Omega$ is the dimensionless volume (number of total lattice sites); $\Delta^{gel}_i$ is the free energy change of species $i$ upon associating to the gel; and $N^{gel}_i$ is number of species $i$ in the gel~\cite{flory1942thermodynamics,flory1953principles,tanaka1989}. Overall the system remains electroneutral, $N_{a} = N_{c_1}+N_{c_2}$, but this does not mean that the sol and gel phases need to independently be electroneutral. 
The number of lattice sites occupied by an alkali cation is taken to be one, with the number of lattice sites occupied by an IL cation $\xi_{c_1}$ and an anion occupies $\xi_a$ lattice sites (note this choice is arbitrary and does not affect the result). The total number of lattice sites, $\Omega$, is given by
\begin{equation}
\Omega = \xi_{c_1}N_{c_1}+\sum_{lm}(l + \xi_a m)N_{lm} + N_{c_2}^{gel} + \xi_aN_{a}^{gel}.
\end{equation}
The volume fraction of a cluster of rank $lm$ is given by
\begin{equation}
    \phi_{lm} = (l + \xi_a m)\tilde{c}_{lm},
\end{equation}
where the dimensionless concentration  ($\#$ per lattice site) of clusters is $\tilde{c}_{lm} = N_{lm}/\Omega$. The dimensionless concentration of all other species are analogously defined. The volume fraction of alkali cations in the sol phase is 
\begin{equation}
    \phi_{c_2}^{sol} = \sum_{lm}l\tilde{c}_{lm},
\end{equation}
and for anions
\begin{equation}
    \phi_a^{sol} = \sum_{lm}\xi_am\tilde{c}_{lm}.
\end{equation}
For IL cations the volume fraction is $\phi_{c_1}=\phi_{c_1}^{sol} = \xi_{c_1}\tilde{c}_{c_1}$. The volume fraction of alkali cations in the gel is simply $\phi_{c_2}^{gel} = \tilde{c}^{gel}_{c_2}$, and analogously the volume fraction of anions $\phi_a^{gel} = \xi_a\tilde{c}^{gel}_{a}$. There are also individual relations for the conservation of each species~\cite{mceldrew2020theory}. 

In experiments and simulations the volume fractions of all the species are specified by prescribing the mole fraction of alkali metal salt, $x$, that is doping the IL (at fixed temperature and pressure, $x$ is the only degree of freedom needed to define the system intensively). Each of the species volume fractions (IL cations, alkali cations, and anions) can be written explicitly in terms of $x$
\begin{align}
    \phi_{c_1}=\frac{\xi_{c_1}(1-x)}{\xi_a+(1-x)\xi_{c_1}+x},
    \label{eq:phic1}
\end{align}
\begin{align}
    \phi_{c_2}=\frac{x}{\xi_a+(1-x)\xi_{c_1}+x},
    \label{eq:c_2}
\end{align}
and
\begin{align}
    \phi_{a}=\frac{\xi_a}{\xi_a+(1-x)\xi_{c_1}+x}.
    \label{eq:phia}
\end{align}
The chemical potential of a rank $lm$ cluster can be determined by differentiating the free energy with respect to $N_{lm}$
\begin{align}
\beta \mu_{lm}&=\ln \phi_{lm}+\Delta_{lm} + 1 - (l + \xi_a m) \tilde{c}_{tot} \nonumber \\
&+ \beta \chi \phi_{c_1}\left[(f_am-m-l+1) - (l + \xi_a m)((f_a-1)\tilde{c}_{c_2}-\tilde{c}_a+\tilde{c}_{tot})\right] \nonumber\\
& +l(1-x)\left( x w_{c_2}^{gel}\frac{\partial \Delta^{gel}_{c_2}}{\partial x}+w_a^{gel}\frac{\partial \Delta^{gel}_{a}}{\partial x}\right),
\label{eq:muclust1}
\end{align}
where $\tilde{c}_{tot}=\tilde{c}_{c_1} + \sum_{lm}\tilde{c}_{lm}$ is the total dimensionless concentration for species ($\#$ per lattice site), and $w_i^{gel}=N_i^{gel}/N_i$ is the fraction of species $i$ in the gel. Note, the explicit form of $\Omega$ has to be used when differentiating. Similarly, the chemical potential of an IL cation can be obtained by differentiating the free energy by $N_{c_1}$
\begin{align}
\beta\mu_{c_1}&=\ln \phi_{c_1}+1-\xi_{c_1}\tilde{c}^{tot} \nonumber \\
&+ \beta \chi\xi_{c_1}\left[(f_a-1)\tilde{c}_{c_2}-\tilde{c}_a+\tilde{c}_{tot}\right]\left( 1-\phi_{c_1} \right) \nonumber \\
&-x^2w_{c_2}^{gel}\frac{\partial \Delta^{gel}_{c_2}}{\partial x}-xw_a^{gel}\frac{\partial \Delta^{gel}_{a}}{\partial x}.
\end{align}

Establishing an equilibrium between all clusters requires the condition:
\begin{align}
\mu_{lm}=l\mu_{10}+m\mu_{01}.
\label{eq:clusteq}
\end{align}
Note the indices 01 and 10 correspond to free alkali cations and anions, respectively. Plugging Eq.~\eqref{eq:muclust1} into Eq.~\eqref{eq:clusteq}, the following mass action law is obtained
\begin{align}
\phi_{lm}=K_{lm}\phi_{10}^{l}\phi_{01}^{m},
\label{eq:clustp}
\end{align}
where $K_{lm}=\exp\left\{(l+m-1)(1+\chi \phi_{c_1})-\Delta_{lm}  \right\}$ is the equilibrium constant and $\Delta_{lm}$ is the free energy of formation for rank $lm$ clusters from free alkali cations and free anions. There are three major contributions to $\Delta_{lm}$: 1) combinatorial entropy, 2) binding energy and 3) configurational entropy. 

The combinatorial contribution is given by
\begin{align}
\Delta_{lm}^{comb}=- \log f_{c_2}^lf_{a}^mW_{lm},
\end{align}
where $W_{lm}$ is the enumeration of the ways that a rank $lm$ cluster can be formed. For Cayley tree clusters (no intra-cluster loops), Stockmayer~\cite{stockmayer1952molecular} determined the exact expression to be 
\begin{align}
W_{lm}=\frac{(f_{c_2}l-l)!(f_{a}m-m)!}{l!m!(f_{c_2}l-l-m+1)!(f_{a}m-m-l+1)!}.
\end{align}

The binding energy of a cluster can be approximated as the energy of a single association between an alkali cation and an anion, $\Delta u_{c_2a}$, and the total number of associations in a cluster, which for Cayley tree clusters with no intra-cluster loops is $l+m-1$. Thus, the binding energy contribution is
\begin{align}
\Delta_{lm}^{bind}=\beta (l+m-1)\Delta U^{bind}_{c_2a}.
\label{eq:bind}
\end{align}
We note that the binding energy $\Delta U^{bind}_{c_2a}$ is largely electrostatic in nature and assumed to be a constant for a given mole fraction of alkali metal salt.

The configurational contribution determines the entropy of placing a rank $lm$ cluster on a lattice with coordination number $Z$. Flory's expression for the so-called entropy of disorientation used in lattice fluid  theory~\cite{flory1942thermodynamics,flory1953principles,tanaka1989,tanaka1999} is employed
\begin{align}
\Delta_{lm}^{conf} = -\ln\left(\frac{l+\xi_am}{\xi_a^m}\right) -(l+m-1)\ln\left(\frac{[Z-1]^{2}}{Ze}\right).
\end{align}
One can go beyond this entropic term and also account for flexibility of associations. For further details, see Ref.~\citenum{mceldrew2020corr}. However, the inclusion of such physics, would only become necessary in modelling the temperature dependence of ion association, which we relegate to later investigations.

Thus, in total, $\Delta_{lm}=\Delta^{comb}_{lm}+\Delta^{bind}_{lm}+\Delta^{conf}_{lm}$, which can be plugged into Eq.~\eqref{eq:clustp} to obtain the thermodynamically consistent cluster distribution
\begin{align}
\tilde{c}_{lm}=\frac{W_{lm}}{\lambda}  \left(\lambda f_{c_{2}}\phi_{10}\right)^l \left(\lambda f_a \phi_{01}/\xi_a\right)^m,
\label{eq:clust}
\end{align}
where $\lambda$ is the ionic association constant given by
\begin{align}
\lambda=\frac{[Z-1]^{2}}{Z}\exp \left\{\beta\left(-\Delta U^{bind}_{c_2a} + \chi \phi_{c_1}\right)\right\}.
\label{eq:L}
\end{align}
Note, it is convenient to factor $\lambda$ as $\lambda=\lambda_0\exp \left\{\chi \phi_{c_1}\right\}$, because $\lambda_0$ is independent of $x$ and $\exp \left\{\beta \chi \phi_{c_1}\right\}$ is not [$\phi_{c_1}=\phi_{c_1}(x)$ via Eq.~\eqref{eq:phic1}]. Furthermore, we will refer to $\lambda_0=\frac{[Z-1]^{2}}{Z}\exp \left\{-\beta\Delta U^{bind}_{c_2a}\right\}$ as the ``bare" association constant, because it is the association constant of the alkali cation and the anion in the absence of any interactions with the IL cation. It is clear from Eq.~\eqref{eq:L} how the IL cation affects the ion association between the alkali cation and anion. Favorable interaction between the IL cation and the anion (negative $\chi$), lowers the ion association constant, and reduces the affinity for association between alkali cations and the anion. 

In Eq.~\eqref{eq:clust}, $\tilde{c}_{lm}$ is written in terms of the volume fraction of free alkali cations ($\phi_{10}$) and IL anions ($\phi_{01}$). However, $\phi_{10}$ and $\phi_{01}$ are, in principle, experimentally inaccessible. Instead, it is natural to express the cluster distribution in terms of the overall volume fractions of each species, $\phi_i$, which is an experimentally and computationally controllable parameter through the mole fraction of the alkali metal salt [see Eqs.~\eqref{eq:c_2}-\eqref{eq:phia}]. This connection is established by introducing ion association probabilities, $p_{ij}$, which is the probability that an association site of species $i$ is bound to species $j$, where $i$ and $j$ are either the alkali cation ($c_{2}$) or the IL anion ($a$). Therefore, the volume fraction of free alkali cations can be written as $\phi_{10} = \phi_{c_{2}}(1 - p_{c_{2}a})^{f_{c_{2}}}$ and free anions as $\phi_{01} = \phi_a(1 - p_{ac_{2}})^{f_a}$. 

The association probabilities can be determined through the conservation of associations and a mass action law between open and occupied association sites. The conservation of associations is given by
\begin{align}
p_{c_{2}a}\psi_{c_2}=p_{ac_{2}}\psi_a,
\label{eq:cons}
\end{align} 
where $\psi_{c_2}=f_{c_2}\phi_{c_2}$ and and $\psi_{a}=f_{a}\phi_{a}/\xi_a$ are the number of alkali cation  and anions association sites per lattice site, respectively. The mass action law between open and occupied association sites is
\begin{align}
\lambda \zeta = \dfrac{p_{c_2a}p_{ac_2}}{(1-p_{c_2a})(1-p_{ac_2})},
\label{eq:MAL}
\end{align}
where $\zeta=\psi_a p_{ac_2}=\psi_{c_2}p_{c_2a}$ is dimensionless concentration of associations (per lattice site). The definition of $\lambda$ as the ionic \emph{association constant} becomes clear from its appearance in the association mass action law [Eq.~\eqref{eq:MAL}]. It sets the equilibrium for association sites to be occupied or open. Equations \eqref{eq:cons} and \eqref{eq:MAL} permit an explicit solution for the probabilities $p_{c_2a}$ and $p_ac_2$ in terms of overall species volume fractions
\begin{align}
\psi_a p_{ac_2}=\psi_{c_2}p_{c_2a}=\frac{1 + \lambda (\psi_a + \psi_{c_2}) - \sqrt{\left[1 + \lambda (\psi_a + \psi_{c_2})\right]^2-4 \lambda^2 \psi_a \psi_{c_2}} }{2 \lambda}.
\label{eq:p}
\end{align}
Note, we may substitute the definitions of $\phi_{10}$ and $\phi_{01}$, as well as Eq.~\eqref{eq:MAL} into Eq.~\eqref{eq:clust}, to obtain an expression for the cluster distribution explicitly in terms of the association probabilities:
\begin{align}
\tilde{c}_{lm}=\mathcal{K}W_{lm}  \left(\frac{p_{ac_2}}{1-p_{ac_2}}(1-p_{c_2a})^{f_{c_2}-1}\right)^l \left(\frac{p_{c_2a}}{1-p_{c_2a}}(1-p_{ac_2})^{f_{a}-1}\right)^m,
\label{eq:clust2}
\end{align}
where $\mathcal{K}=\psi_{c_2}(1-p_{ac_2})(1-p_{c_2a})/p_{ac_2}=\psi_{a}(1-p_{ac_2})(1-p_{c_2a})/p_{c_2a}$. Thus, it can be seen plainly that the association probabilities defined in Eq.~\eqref{eq:p} can be substituting into Eq.~\eqref{eq:clust2} to obtain the full distribution of ion clusters in the electrolyte explicitly as a function of the overall species volume fractions, as well as the model parameters ($f_{i}$,$\xi_{i}$, $\lambda_0$,$\chi$). As we will show later, when $\psi_a$ is very different than $\psi_{c_2}$ our model tends to predicts very asymmetric ($l \neq m$) cluster distributions, much in line with observations from molecular simulations \cite{li2012li+,molinari2019general,molinari2019transport,molinari2020chelation,lourenco2021theoretical}.

The association probabilities defined in Eq.~\eqref{eq:p} can be linked directly to the average coordination number of alkali cations by anions ($f_{c_2}p_{c_2a}$), as well the coordination number of anions by alkali cations ($f_ap_{ac_2}$). Such quantities are somewhat accessible experimentally, as anions that are coordinated to alkali cations display a shift in Raman spectroscopy bands, and the relative fractions coordinated and free anions can be reliably measured \cite{lassegues2009spectroscopic,menne2014lithium,pitawala2015coordination,reber2021anion}. Such data can be analyzed and understood within our framework. 

\subsection{Ion Network Formation}

A key prediction of our theory is the formation of a percolating ionic network, when $x$ exceeds a critical threshold, $x^*$. This can be clearly seen by observing the mathematical form of the weight-averaged cluster size (or degree of aggregation), $\bar{n}$, which can be expressed analytically as 
\begin{align}
    \bar{n} &= \frac{\sum_{lm}(l+m)^2c_{lm}}{\sum_{lm}(l+m)c_{lm}} \nonumber \\
    &=
    1+\left(\frac{f_{c_2}}{p_{ac_2}}+\frac{f_{a}}{p_{c_2a}}\right)^{-1}\left[ \frac{(f_{c_2}-1)(p_{c_2a})^2p_{ac_2}+(f_{a}-1)(p_{ac_2})^2p_{c_2a}+2p_{c_2a}p_{ac_2}}{1-(f_{c_2}-1)(f_{a}-1)p_{c_2a}p_{ac_2}} \right].
    \label{eq:barn}
\end{align}
The weight-averaged degree of aggregation diverges when $p^*_{c_2a}p^*_{ac_2} = 1/(f_{c_2}-1)/(f_a-1)$ which corresponds to the appearance of a percolating ion network in the electrolyte. In polymer physics, this critical condition has been historically termed the gel point, so the percolating ion network will be referred to as the  ``gel" interchangeably.  It is clear that the condition for ion network formation can be satisfied if the functionalities of both ions are greater than 1~\cite{mceldrew2020theory}. Furthermore, if either the anion or alkali cation have functionalities greater than 2, then the ion network can form when the association probabilities are less than 1. An implicit expression for the gel point can be readily obtained by plugging Eq.~\eqref{eq:p} into the percolation criterion ($p_{c_2a}p_{ac_2} = 1/(f_{c_2}-1)/(f_a-1)$)
\begin{align}
    \lambda^* = \left[\frac{1+(f_a-1) (f_{c_2}-1)}{
     \sqrt{(f_a-1) (f_{c_2}-1)/ \psi_a \psi_{c_2}}}-\psi_{c_2} -\psi_a\right]^{-1}.
     \label{eq:gp}
\end{align}
Recall, that $\psi_a$ and $\psi_{c_2}$ can be written explcitly in terms of $x$ via Eqs. \eqref{eq:c_2} and \eqref{eq:phia}. Thus, Eq.~\eqref{eq:gp} should be regarded as an implicit equation for the critical alkali metal salt fraction, $x^*$, as well. 

As we have mentioned, percolating ion networks have been observed in molecular simulations in super-concentrated electrolytes, including  the salt-in-ionic liquid systems studied in here. In principle, solid-like behavior can be observed for systems in which percolating molecular networks are present.  If the timescale of associations are comparable to the molecular relaxation times, then percolation can induce solid-like character in those relaxation processes. In Refs.~\citenum{li2012li+,molinari2019transport,molinari2020chelation}, the anionic coordination shell of salt-in-ionic liquid electrolytes largely persists for nanosecond timescales. Such timescales are expected to be significantly long-lived, so as to be manifested experimentally.  Indeed, in recent experimental work~\cite{reber2021anion}, Reber \textit{et al.} observed that sodium salt in ionic liquid systems tend to solidify or form a gel when the sodium salt fraction exceeds a certain threshold, and this gelation point seems to be in rough agreement with the salt fraction at which a percolating ion network first appears in molecular simulations of the same system~\cite{molinari2019transport}. Thus, there is reason to suspect that the formation of ion network can trigger a structural arrest (gelation) of the salt-in-ionic liquid system. 

Once the critical percolation threshold has been reached, the volume fractions of alkali cations and anions in the gel (ion network), $\phi_{c_2}^{gel}$ and the sol (electrolyte species excluded from the ion network) $\phi_a^{sol}$ must be determined. In order to do this, Flory's treatment of the post-gel regime is employed, in which the volume fraction of free ions can be written equivalently in terms of overall association probabilities, $p_{ij}$, and association probabilities taking into account only the species residing in the sol, $p^{sol}_{ij}$
\begin{align}
\phi_{c_2}(1-p_{c_2a})^{f_{c_2}}=\phi_{c_2}^{sol}(1-p^{sol}_{c_2a})^{f_{c_2}}
\label{eq:gel1}
\end{align}
\begin{align}
\phi_a(1-p_{ac_2})^{f_{a}}=\phi_a^{sol}(1-p^{sol}_{ac_2})^{f_{a}}
\label{eq:gel2}
\end{align}
Where $\phi_{c_2/a}^{sol}=1-\phi_{c_2/a}^{gel}$ is the volume fraction of alkali cations or anions in the sol. The two unknown $\phi_i^{sol}$ variables, as well as the two unknown sol association probabilities, $p_{ij}^{sol}$, can be determined using Eqs.~\eqref{eq:gel1}-\eqref{eq:gel2} in addition to Eqs.~\eqref{eq:cons}-\eqref{eq:MAL}, however in this case we use sol-specific quantities. Note that prior to the critical gel concentration, there is a trivial solution that $p_{ij}=p^{sol}_{ij}$ and $\phi_i=\phi^{sol}_i$, yielding a gel volume fraction of $\phi_i^{gel}=0$. However, beyond the gel point, there is a non-trivial solution that yields $\phi_i^{gel}>0$, marking the emergence of the percolating ion network with finite volume fraction. One seemingly general trend is that beyond the gel point, as the gel increases in volume fraction, the sol association probabilities tend to decrease. This means that the ions excluded from the gel tend to be less associated and thus more free. As we will show, this has a major implication on the observed effective charge of alkali cations in the electrolyte; for $x>x^*$, indeed ion associations increase overall, but \emph{alkali cations actually become less bound to anions in the sol}.

\subsection{Effective Charge of Alkali Cations}

It has been observed in both experimentally~\cite{gouverneur2018negative} and from simulations~\cite{molinari2019general,molinari2019transport,molinari2020chelation} that alkali cations display anomalous transport behavior when dissolved in ILs. In particular, lithium and sodium ions were observed to have negative transference numbers when dissolved in ILs at low to moderate mole fraction, which were then reversed to positive values at higher mole fractions~\cite{gouverneur2018negative,molinari2019general}. This behaviour was rationalized in terms of the ``effective" charge of alkali cations ions. Essentially, at low mole fractions of alkali metal salt, each alkali cation is greatly outnumbered by anions, and thus, every cation tends to be surrounded by a coordinating shell of anions. This is important because the alkali cations can diffuse within the electrolytes via a ``vehicular mechanism" with its coordinating anions. In this way, the alkali cation plus its shell of coordinating anions has a \emph{net negative effective charge} resulting in the observed negative transference numbers. Such negative transference numbers often result in large concentration gradients during the operation of energy storage devices, which translates into increased internal resistance for the device. Of course, it is possible that the ions diffuse within the electrolyte via non-vehicular mechanisms (especially at high alkali metal salt fraction~\cite{chen2016elucidation}), such as activated hopping processes, but such processes cannot explain the observations of negative cation transference.

\newif\ifshow\showfalse
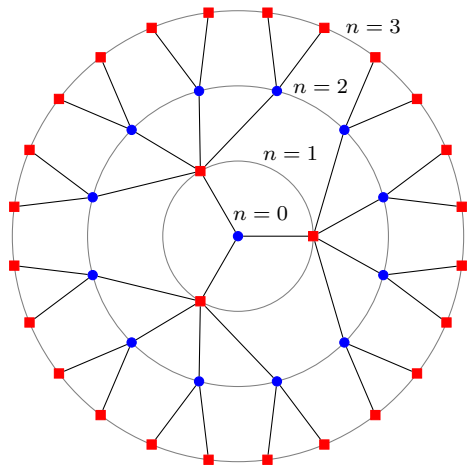
\begin{figure}
\begin{tikzpicture}
\node[] at (0.3,0.3) {\scriptsize $n = 0$};
\node[] at (0.7,1.1) {\scriptsize $n = 1$};
\node[] at (1.1,2) {\scriptsize $n = 2$};
\node[] at (1.8,2.8) {\scriptsize $n = 3$};

\def\j{1}\def\dlast{1}\def\tlast{0}
\foreach \i [remember=\i as \ilast] in {0,...,3}
{
\draw [gray] (0,0) circle (\i cm);
\pgfmathsetmacro\a{360/\j}
\ifodd\i\def\s{}\def\c{red}\def\d{4}\pgfmathsetmacro\t{\tlast-1.5*\a}\else\def\c{blue}\def\s{circle}\def\d{2}\pgfmathsetmacro\t{\tlast-2.5*\a}\fi
\ifnum\i=1\pgfmathsetmacro\t{-\a}\fi
\ifnum\i=0\def\d{3}\def\t{0}\fi
\foreach \k [evaluate=\k as \m using {(\k*\a)+\t}, evaluate=\k as \g using {int((floor((\k-1)/\dlast)))}, count=\n from 0 ] in {1,...,\j}
{
\ifshow\def\tempa{n-\i-\g-\n:\k}\else\let\tempa\relax\fi
\node (n-\i-\n) [draw, fill, \c, \s, minimum size=3.5pt, inner sep=0pt, label={[font=\tiny]{\tempa}} ] at (\m:\i cm) {};
\ifnum\i>0 \draw [black] (n-\i-\n) -- (n-\ilast-\g); \fi
}
\pgfmathsetmacro\j { \j*\d }
\global\let\j\j
\global\let\dlast\d
\pgfmathsetmacro\tlast{(\i==1) ? 0 : (\t+\a) }
\global\let\tlast\tlast
}
\end{tikzpicture}
\caption{Schematic of the shell structure of the alternating Cayley tree clusters where cations (red) are shown to have a functionality of three, and anions (blue) are shown to have a functionality of 4. Additionally, the cluster is truncated to the third shell, and  all of the nodes are shown to be occupied by ions, though in principle the nodes will  only be occupied in accordance with a prescribed association probability.}
\label{fig:shell}
\end{figure}

As we have mentioned the negative effect cation transference phenomenon does not remain at high alkali metal salt mole fractions. A principal reason for this is the appearance of the percolating ion network of lithium ions and anions. The ion network itself is macroscopic and cannot diffuse appreciably as a whole, and thus will not conduct any ionic current. However, species in the sol, can and will be the principal conductors of ionic current in the system. As the network increases in size, the species excluded from the network (though reduced in number) actually become more free. Thus, the effective charge of lithium increases to positive values, and eventually reaches a value of roughly one. Although the phenomenology of this unexpected observation was well-explained in Ref.~\citenum{molinari2019general}, a precise theoretical description of this phenomenon has yet to be put forth. 

Our theory of ion association and network formation is well-equipped to provide such a description. Mathematically, we can compute the effective charge of an alkali cation is via the ``shell method". In this method, we consider an average cluster with a central alkali cation, and compute the average charge of successive shells surrounding the central cation. This is visualized in Fig.~\ref{fig:shell}. A ``zero shell" approximation only considers the charge of the central alkali metal cation, i.e. $q^{eff,0}_{c_2}=1$. 

A ``single shell" approximation can be computed straight away by considering the average charge of the zeroth and first shell. The first shell strictly contains anions that are associated to the central cation, and therefore, its charge will be $-f_{c_2}p_{c_2a}$. Thus, in the single shell approximation, the effective charge of the alkali metal cations is $q_{c_2}^{eff,1}=1-f_{c_2}p^{sol}_{c_2a}$. Note, that we use ``sol" probabilities, as seen in Eqs.~\eqref{eq:gel1} and \eqref{eq:gel2}, because we only want the effective charge to include ion clusters that can contribute to ion conduction, and the network cannot diffuse due to its macroscopic length scale~\cite{mceldrew2020corr}. This assumption is justified since, according to the Stokes-Einstein relation the diffusion coefficient scales inversely of the size of the species, the diffusion coefficient of the ionic network should be vanishingly small.

The general case for $N$ shells is given by
\begin{equation}
q_{c_2}^{eff,N}=1+\sum_{n=1}^{N}q_n,
\label{eq:gen_eff}
\end{equation}
where $q_n$ is the average charge of the $n$th shell of an ionic cluster with a central alkali cation. Note that while it might be tempting to interpret Eq.~\eqref{eq:gen_eff} to suggest the central ion is spatially correlated with the ions in the $N$ shell, this is not an essential condition since, as we shall now show, we only require spatial correlations between ions in neighbouring shells. We can obtain a general formula for $q_n$ by first noticing that the number of nodes in the $n$th shell, $N_n$, in a Cayley tree with a central cation with alternating cationic and anionic nodes (with generally different functionalities) is the following
\[
N_n=f_{c_2}
    \begin{cases} 
      \left(f_{c_2} - 1\right)^{n/2 - 1/2}\left(f_a - 1\right)^{n/2-1/2} & \text{odd } n  \\
      \left(f_{c_2} - 1\right)^{n/2 - 1}\left(f_a - 1\right)^{n/2} & \text{even } n 
   \end{cases}
\]
Similarly, the probability that nodes in the $n$th shell, $P_n$ will be occupied is the following
\[
P_n=p^{sol}_{c_2a}
    \begin{cases} 
      \left(p^{sol}_{c_2a}\right)^{n/2 - 1/2}\left(p^{sol}_{ac_2}\right)^{n/2-1/2} & \text{odd } n  \\
      \left(p^{sol}_{c_2a}\right)^{n/2 - 1}\left(p^{sol}_{ac_2}\right)^{n/2} & \text{even } n 
   \end{cases}
\]
Finally, the nominal charge of ions in the $n$th shell will simply be $(-1)^n$, i.e. alternating between $+$ and $-$ charges. Thus, the average charge of the $n$th shell, $q_n=(-1)^nN_nP_n$, is 
\[
q_n=(-1)^{n}f_{c_2}p^{sol}_{c_2a}
    \begin{cases} 
      \left\{p^{sol}_{c_2a}(f_{c_2} - 1)\right\}^{n/2 - 1/2}\left\{p^{sol}_{ac_2}(f_a - 1)\right\}^{n/2-1/2} & \text{odd } n  \\
      \left\{p^{sol}_{c_2a}(f_{c_2} - 1)\right\}^{n/2 - 1}\left\{p^{sol}_{ac_2}(f_a - 1)\right\}^{n/2} & \text{even } n 
   \end{cases}
\]

If $N$ is odd, then there are an equal number of cationic and anionic shells in the cluster, which will tend to give values of the effective charge that are closer to zero since the paired shells form a generalised ``ion pair" structure reminiscent of overscreening. For values of $N > 1$ that are odd, Eq.~\eqref{eq:gen_eff} can be solved to give
\begin{align}
q^{eff,N}_{c_2}= 1 - \frac{[1 - p^{sol}_{ac_2}(f_a-1)] f_{c_2}p^{sol}_{c_2a}(1 - [(f_{a}-1) (f_{c_2}-1) p^{sol}_{ac_2} p^{sol}_{c_2a}]^{(N+1)/2})}{1 - (f_{a}-1) (f_{c_2}-1) p^{sol}_{ac_2} p^{sol}_{c_2a}}.
\label{eq:qn}
\end{align}
In the case of a finite $N$, the effective charge will never diverge to infinity. Interestingly, this general solution gives the same overall form of equation as the single shell approximation, but there is an \emph{effective} charge of the anions $q^{eff,N}_{c_2} = 1 + \tilde{q}^{eff,N}_{a}f_{c_2}p^{sol}_{c_2a}$. For $N < 3$, one must set $\tilde{q}^{eff,N}_{a} = -1$. One can derive an analogous expression for when $N$ is even, but this is more cumbersome than the odd $N$ expression. 

In the infinite shell limit, Eq.~\eqref{eq:qn} reduces to
\begin{align}
q^{eff,\infty}_{c_2}= 1 - \frac{[1 - p^{sol}_{ac_2}(f_a-1)] f_{c_2}p^{sol}_{c_2a}}{1 - (f_{a}-1) (f_{c_2}-1) p^{sol}_{ac_2} p^{sol}_{c_2a}}.
\label{eq:qinf}
\end{align}
It can be seen clearly that $q^{eff,\infty}_{c_2}$ diverges exactly at the gel point. This divergence is a direct result of the diverging weight-averaged degree of aggregation. Moreover, the excess of anions in relation to alkali cations dictates that clusters will generally be negatively charged. Thus, the diverging cluster sizes will be accompanied by a diverging charge of those clusters. 

Accounting for charge outside the first few shells might not correlate strongly with the transport of alkali cations, however. This is because ion associations have a finite lifetime, and therefore large clusters break apart before they can appreciably diffuse~\cite{mceldrew2020corr}. A similar argument was made in Ref.~\citenum{borodin2020uncharted}; the distance that a a species travels during the residence time of an association with a ligand, should be larger than the size of that ligand, in order for the ion to vehicularly diffuse with the ligand. Thus, the vehicular transport of alkali cations might realistically only occur with a very small number of shells, depending on the residence time of associations. The determination of the precise number of shells to include, would likely require a more precise knowledge of the association dynamics of a given salt-in-IL system, which is beyond the scope of our current study.

Ultimately, the finite lifetime of ion associations means that the effective charge of alkali cations should be more directly correlated to $q_{c_2}^{eff,1}$ than $q_{c_2}^{eff,\infty}$. Continuing with this notion, $q_{c_2}^{eff,1}$ is the charge of the central alkali cation plus the average charge of its first shell. However, $q_{c_2}^{eff,1}$ includes contributions from clusters that extend past the first shell. In principle, the clusters extending further than one shell will not appreciably diffuse, before they break up and transport as smaller clusters. Perhaps, then, a more appropriate approximation for the effective charge of the alkali cation is the average charge of clusters that do not extend past a single shell, i.e. clusters containing a single alkali cation 
\begin{equation}
    q_{c_2}^{eff,1*}= \dfrac{\sum_{1m}(1-m)\tilde{c}_{1m}}{\sum_{1m}\tilde{c}_{1m}},
    \label{eq:q1s}
\end{equation}
which can be written in closed form as 
\begin{equation}
q_{c_2}^{eff,1*}=1-\dfrac{f_{c_2}p^{sol}_{c_2a}(1 - p^{sol}_{ac_2})^{f_a-1}}{1 - p^{sol}_{c_2a} + p^{sol}_{c_2a}(1 - p^{sol}_{ac_2})^{f_a-1}}.
\label{eq:gen_eff_one_s}
\end{equation}

\begin{figure}
\centering
\begin{minipage}[b]{0.45\textwidth}
\centering 
\includegraphics[width=\textwidth]{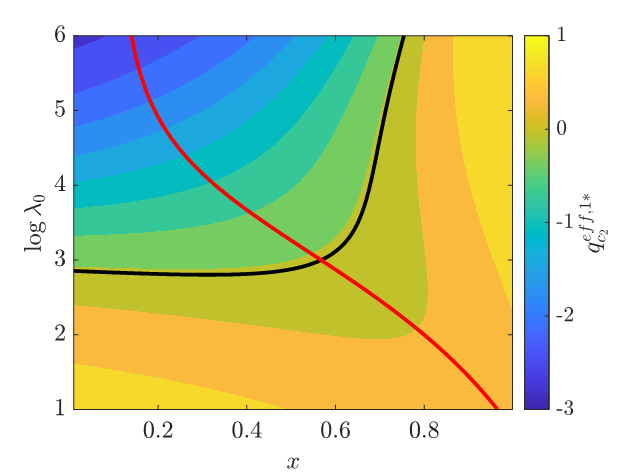}
\end{minipage}
\begin{minipage}[b]{0.45\textwidth}
\centering 
\includegraphics[width=\textwidth]{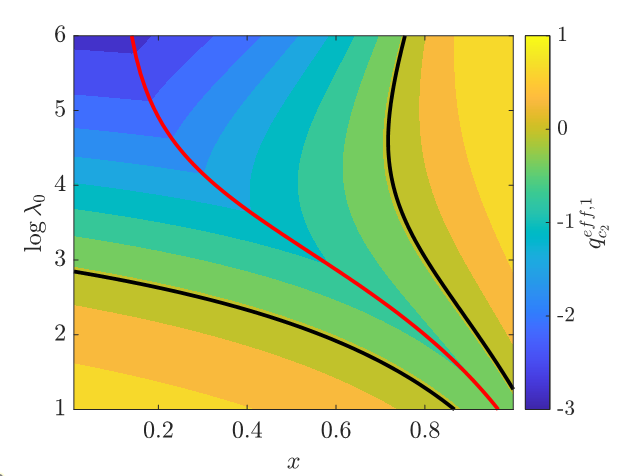}
\end{minipage}
\centering
\begin{minipage}[b]{0.45\textwidth}
\centering 
\includegraphics[width=\textwidth]{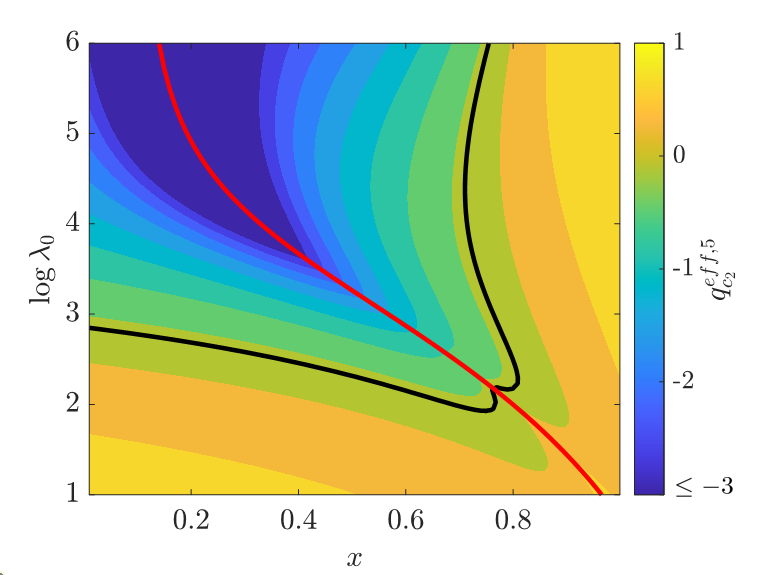}
\end{minipage}
\centering
\begin{minipage}[b]{0.45\textwidth}
\centering 
\includegraphics[width=\textwidth]{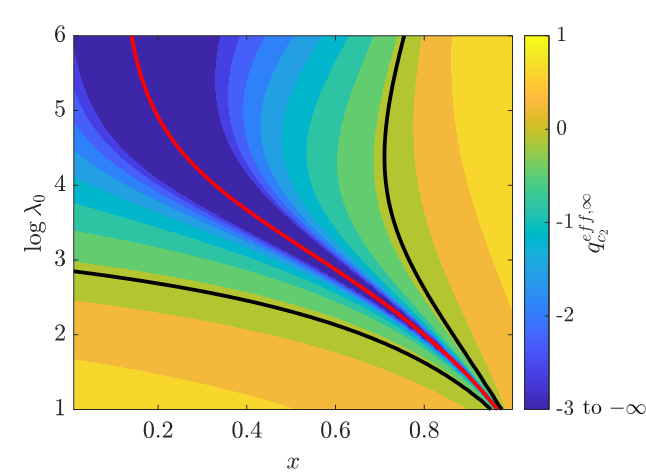}
\end{minipage}

\caption{Effective charges in different approximations: $q^{eff,1*}_{c_2}$ (top left), $q^{eff,1}_{c_2}$ (top right), $q^{eff,5}_{c_2}$ (bottom left), and $q^{eff\infty}_{c_2}$ (bottom right) as indicated in the respective colour bar, as a function of mole fraction of alkali metal salt and bare ionic association ($\lambda_0$), for a fixed regular solution interaction $\chi=-3~k_BT$. Note, in the bottom left and bottom right panels, the color bar remains dark navy blue for all charges less than or equal to negative 3. In each panel, we draw the critical gel boundary (red line), as well as the zero contour line for the effective charge of the alkali-cation (black line).}
\label{fig:q_shell}
\end{figure}

In order to demonstrate the trends of each effective charge formula, we plot color maps of $q_{c_2}^{eff,1*}$, $q_{c_2}^{eff,1}$, $q_{c_2}^{eff,5}$, and $q_{c_2}^{eff,\infty}$ in Fig.~\ref{fig:q_shell} as functions of $x$ and $\lambda_0$ with a fixed value of $\chi=-3~k_BT$. We see generally that the low-$x$ and large-$\lambda_0$ region of space cause the effective charge of alkali cations to take on negative values no matter what approximation for $q_{c_2}^{eff}$ is employed. In this region (low $x$ and large $\lambda_0$), alkali cation associations with anions is very favorable and there will be significant vehicular transport because most ions are not bound up in a percolating ionic network. For low $x$ and low $\lambda_0$, all approximations predict net positive effective charges of the alkali cation, as would be expected from weakly associating ions. 

For large $x$ and low $\lambda_0$, however, there are some discrepancies between the predictions of each approximation. The shell approximation tends to predict the formation of net negative alkali cations, while the $q_{c_2}^{eff,1*}$ method clearly predicts net positive charges, as is evident from the neutral line (black curve). This occurs because the net charges predicted with the shell method go beyond what can be vehicularly transported with a single alkali cation. Moreover, this method causes a divergence in the net charge of the alkali cation when infinite shells are accounted for close to the gel point (red curve), as the gel (of infinite size) does not comprise of equal numbers of cations and anions. For large $x$ and large $\lambda_0$, all approximations predict net positive charges of alkali cations, as the gel phase has incorporated all large clusters, leaving only free alkali cations.

\begin{figure}
\centering
\begin{minipage}[b]{0.95\textwidth}
\centering 
\includegraphics[width=\textwidth]{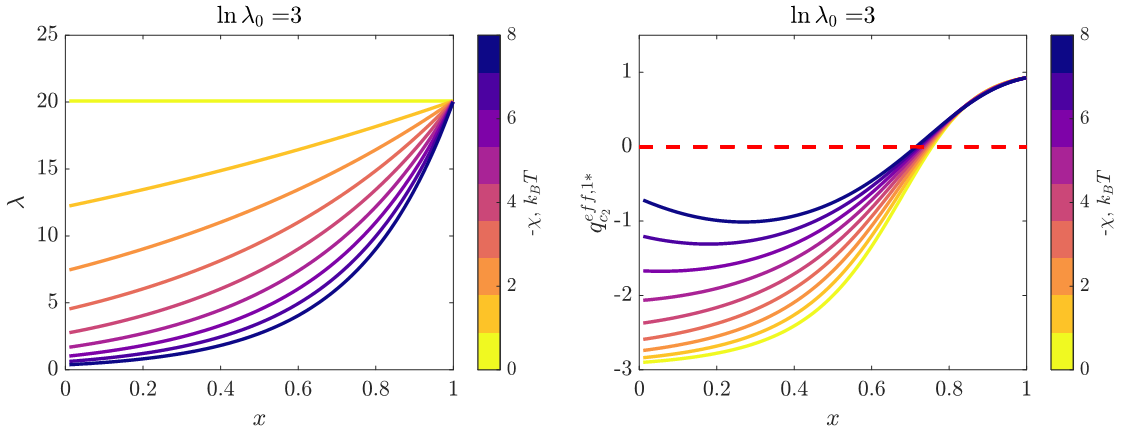}
\end{minipage}
\caption{(left) - Total ionic association constant as a function of mole fraction of alkali metal salt for various regular solution interactions for a fixed bare ionic association constant. (right) - Effective charge of the alkali cation, computed using $q_{c_2}^{eff,1*}$, for the same set of parameters as the left.}
\label{fig:q_1s}
\end{figure}

The primary parameter controlling the association equilibrium in our system is $\lambda$, as seen in Eq.~\eqref{eq:L},which is principally a function of the components in our electrolyte, as well as temperature~\cite{mceldrew2020theory,mceldrew2020corr,mceldrew2021ion}. Thus, a primary function of our model is to provide a framework in understanding how the choices of electrolyte components affect $\lambda$ and ion association overall. This can be seen more directly by recalling our factoring of $\lambda$ as $\lambda=\lambda_0\exp\left(\beta \chi\phi_{c_2}\right)$, where $\lambda_0=\frac{(z-1)^2}{z}\exp\left(-\beta\Delta U^{bind}_{c_2a}\right)$. In principal, $\lambda_0$ and $\chi$ can be tuned independently modelling different alkali metal and IL cations, respectively, for a given shared anion. However, if we were to modify the anion, then $\lambda_0$ and $\chi$ would both have to change accordingly. In contrast to $\lambda_0$, $\exp\left(\beta \chi\phi_{c_2}\right)$ will be a strong function of $x$. Thus, depending on the magnitude of $\chi$, $\lambda$ can change drastically as the alkali metal salt fraction increases. We can see in Fig.~\ref{fig:q_1s} that for favorable IL cation--anion interactions (negative $\chi$ for a given $\lambda_0$), $\lambda$ increases drastically as a function of $x$, when the fraction of IL cation is diminished. Furthermore, we can see the effect of the IL cation--anion interaction, on the effective charge ($q_{c_2}^{eff,1*}$) of the alkali cation in the right panel of Fig.~\ref{fig:q_1s}. Not surprisingly, the IL cation--anion interaction has the largest effect when $\chi$ is the most negative and at low alkali metal salt fraction, where the volume fraction of the IL cation is largest. In this scenario, the IL cation is able to compete with alkali cations ions for interactions with the anions, and the coordination shell of the alkali cation by anions is reduced, even leading to non-monotonic behavior in $q_{c_2}^{eff,1*}$, as a result. 

\section{Results and Discussion}

\begin{figure}
\centering
\begin{minipage}[b]{0.9\textwidth}
\centering 
\includegraphics[width=\textwidth]{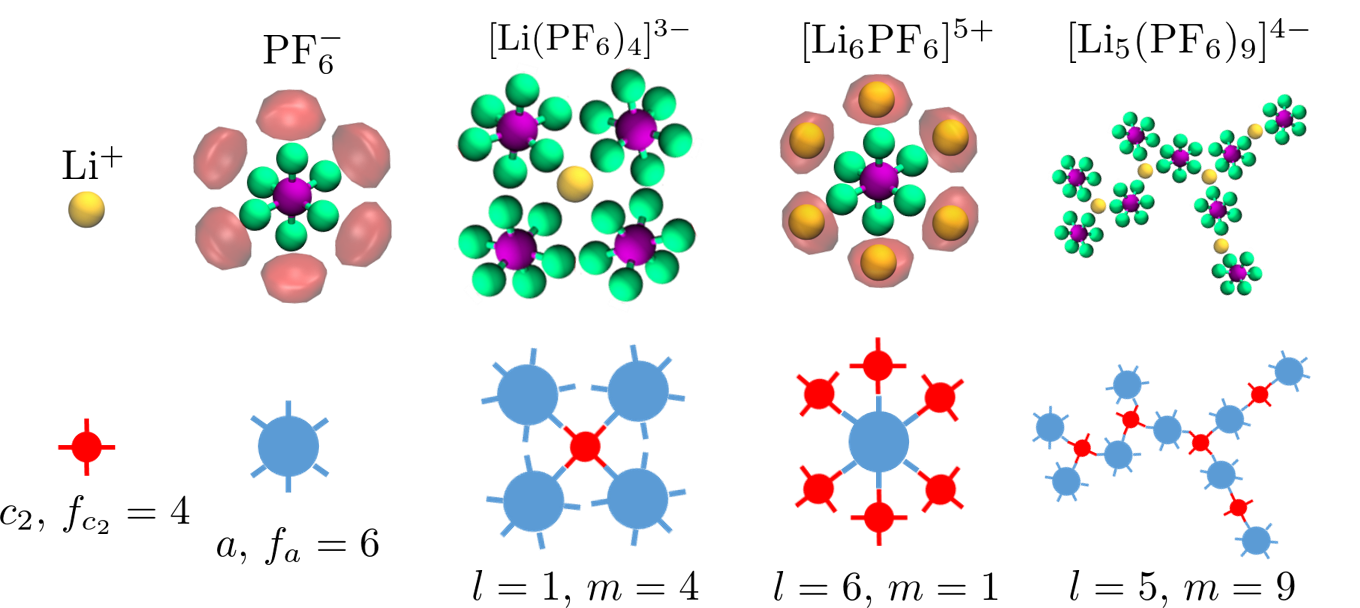}
\end{minipage}
\caption{Schematic of the functionality of lithium cations and PF$_6^-$ anions, and some example clusters they can form. For PF$_6^-$ centred clusters, ``hot-spots" of the lithium cations are shown with red iso-surfaces. The bottom row show the corresponding clusters and cluster numbers in our thermodynamic  ion cluster theory.}
\label{fig:model_MD}
\end{figure}

We now proceed to extract all of the required parameters for our theory from the molecular simulations of Ref.~\citenum{molinari2019general} of emimPF$_6$/LiPF$_6$ and emimBF$_4$/LiBF$_4$ salt-in-ionic liquid systems. The main parameters that need to be determined are the functionalities of each ion ($f_{c_2}$ and $f_a$), the bare association constant ($\lambda_0$) and the regular solution interaction parameter ($\chi$). 

The full set of cluster distributions were computed from molecular dynamics simulations in Ref.~\cite{molinari2019general}, for details as to the precise definitions for associations, including the computed radial distribution functions for emimPF$_6$/LiPF$_6$ and emimBF$_4$/LiBF$_4$ systems, we direct the read to Ref.~\cite{molinari2019general}. In brief, the associations can be determined from MD simulations via a distance threshold value corresponding to the first peak of the Li-F correlation functions, in which the F atom either belongs to PF$_6^-$ or BF$_4^-$.

As noted in Ref.~\citenum{mceldrew2020corr}, ion functionalities for molecular ions can be essentially visualized by the spatial distribution functions (SDFs) of counter-ion around the central ion of interest. We can see this explicitly in Fig.~\ref{fig:model_MD}, where we have drawn a PF$_6^-$ with surrounding iso-density surfaces, showing regions where Li$^+$ is most likely to associate. We immediately see that there are 6 distinct localized regions that Li$^+$ tends to prefer around PF$_6^-$. In this case, we identify these regions as the association sites of PF$_6^-$ and define its functionality to be 6. In a similar fashion, there are 4 distinct localized regions that Li$^+$ tends to prefer around BF$_4^-$, and thus we infer the functionality of BF$_4^-$ to be 4. 

\begin{figure}
\centering
\begin{minipage}[b]{\textwidth}
\centering 
\includegraphics[width=\textwidth]{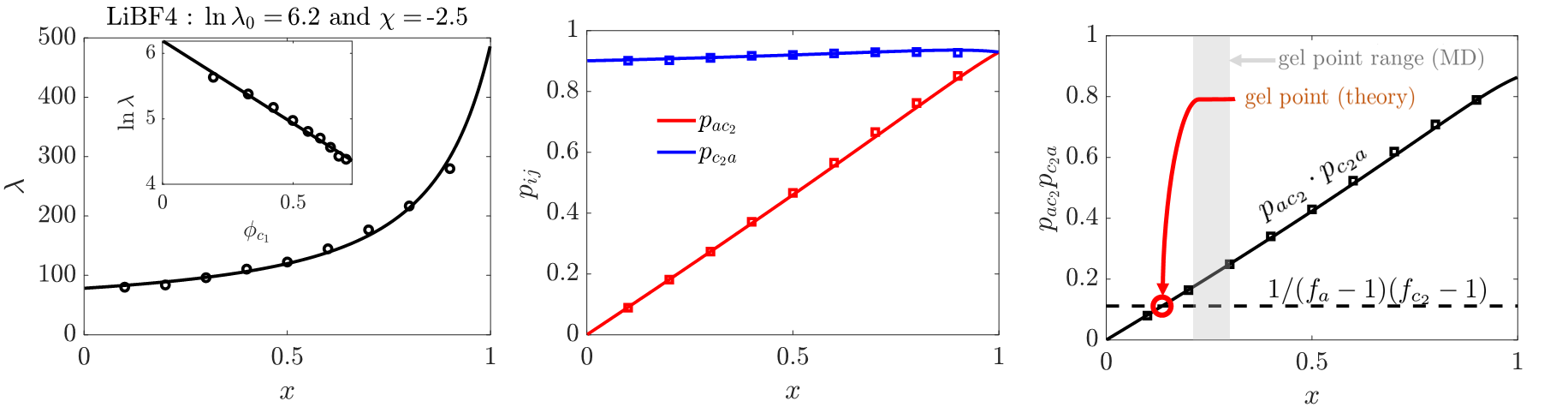}
\end{minipage}
\centering
\begin{minipage}[b]{\textwidth}
\centering 
\includegraphics[width=\textwidth]{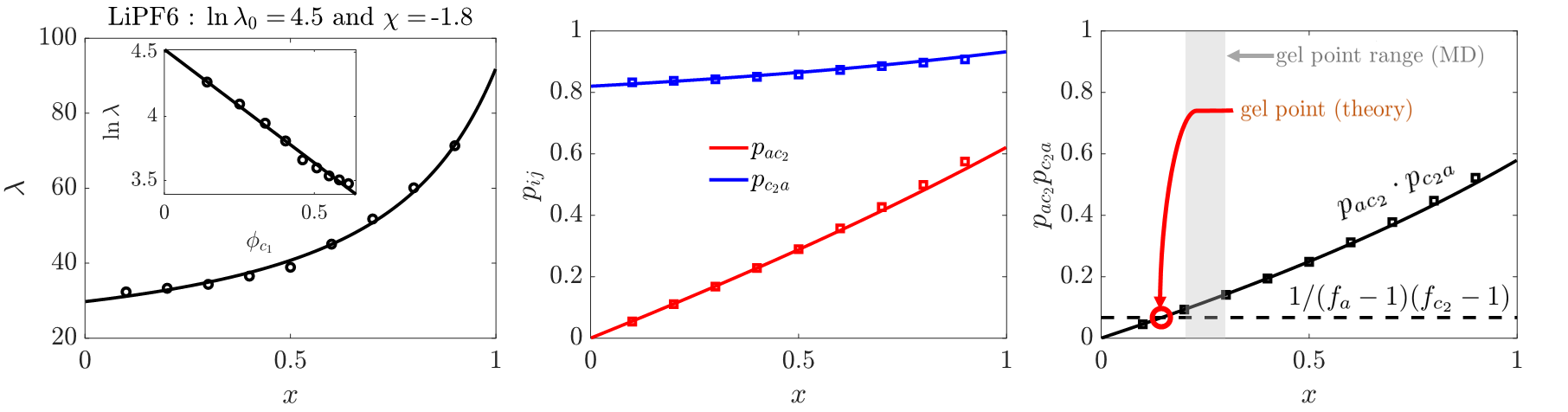}
\end{minipage}
\caption{ The association constant ($\lambda$, left panel), association probabilities ($p_{c_2a}$ and $p_{ac_2}$, middle panel), and percolation threshold ($p_{c_2a} \cdot p_{ac_2}$, right panel) as determined from MD simulations (squares) and theory (solid curves) for both emim/LiBF$_4$ (top panels) and emim/LiPF$_6$ (bottom panels) systems. Here the association constants were fitted according to the function $\lambda=\lambda_0\exp\left\{ \beta \chi \phi_{c_1}\right\}$, and the fitted values for each system are shown in the left panel. }
\label{fig:emim_libf4_lipf6}
\end{figure}

For atomic species, we can not use the convenient SDF visualizations to specify ion functionality. Thus, for lithium, we must look at the distribution of associations that lithium ions tend to make in the simulation. The majority of lithium ions make less than four associations to anions in both Emim/LiBF$_4$ and Emim/LiPF$_6$ systems. Thus, Li$^+$ is best approximated with an ion functionality of 4. This means that Li$^+$ is modelled with the capacity to host 4 anions for coordination. However, there are rare cases observed in MD simulations, which become more probable at higher concentrations of lithium salt, where Li$^+$ can host 5 anions (one excess association). The physical picture is that Li$^+$ can host the first 4 anions, but the fifth association is considerably more unlikely and much less favorable energetically than each of the first 4 anionic associations~\cite{mceldrew2021ion}. There is also a similar phenomenon that occurs with anions: at high concentrations of lithium salt, anions are able to host one more lithium ion than the number of SDF hot spots would suggest. Again, the physical picture is that, this excess association is considerably more energetically unfavorable than the preceding associations, and therefore we do not take these explicitly into account. Thus, we have a direct connection between MD simulations and our thermodynamic model. This connection is further clarified by Fig.~\ref{fig:model_MD}.

Given that the total number of associations are known for a given snapshot (from simulation data in Ref.~\cite{molinari2019general}), and the ion functionalities are specified, we may compute the average association probabilities from MD simulations: $p^{MD}_{ij}=\langle M_{ij}\rangle/(f_i N_i)$, where $\langle M_{ij}\rangle$ is the average number of associations of type $ij$ in the simulation and $N_i$ is number of molecules of type $i$ in the simulation. With the average association probabilities known, we may simply use the mass action law [Eq.~\eqref{eq:L}] to compute the average association constant, $\lambda$ from simulation. A detailed explanation of this procedure was given in Ref.~\citenum{mceldrew2020corr}. $\lambda$ is a strong function function of $x$, as shown in the left panel of Fig.~\ref{fig:emim_libf4_lipf6}. This x-dependence of $\lambda$, is well modelled by the function: $\lambda=\lambda_0\exp \left\{\chi \phi_{c_1}\right\}$ with fitted values of $\chi$ and $\lambda_0$. Moreover, we see that for both LiBF$_4$ and LiPF$_6$, $\log \lambda_0$ is considerably larger in magnitude than $|\beta \chi \phi_{c_1}|$. For example, $|\chi \phi_{c_1}|$ maxes out at around 1~$k_B T$ for either studied system, whereas $k_BT \log \lambda_0$ is 6.2 for LiBF$_4$ and 4.8$k_BT$ for LiPF$_6$. This implies that the energetics of Li-anion associations are considerably more favorable than that of the interactions between IL cations and anions. Furthermore, the fitted regular solution parameter for the Emim$^+$--BF$_4^-$ ($\chi=-2.5~k_BT$) was actually found to be remarkably close to the association energy of Emim$^+$--BF$_4^-$ ($\Delta U^{bind}_{c_1a}=-2.3~k_BT$) found in Ref.~\citenum{mceldrew2020corr}. This provides some confirmation that our treatment of IL cation--anion interactions via the regular solution parameter was valid. Of course, treating IL cation--anion associations explicitly would be more consistent within our framework, but our regular solution treatment seems to capture the phenomenology without being subjected to the additional mathematical complexity.  

The fitted $\lambda$ function also produces a good fit for the association probabilities, given by Eq.~\eqref{eq:p} in the pre-gel regime and additionally Eqs.~\eqref{eq:gel1} and \eqref{eq:gel2} in the post-gel point regime, as shown in Fig.~\ref{fig:emim_libf4_lipf6}. We can see in the middle panel of Fig.~\ref{fig:emim_libf4_lipf6} that both $p_{c_2a}$ and $p_{ac_2}$ increase monotonically as functions of $x$. However, $p_{ac_2}$ clearly increases much more drastically than $p_{c_2a}$. This is simply because there are more available alkali cations to occupy anion association cites as $x$ increases, resulting in an increasing $p_{ac_2}$. This same notion would actually tend to reduce the $p_{c_2a}$. However, $\lambda$ strongly increases as a function of $x$, and therefore, although there are less anions per alkali cation to occupy $c_2$ association sites, alkali cations and anions associate more vigorously as $x$ increases. 

In the right panel of Fig.~\ref{fig:emim_libf4_lipf6}, we plot the product, $p_{c_2a} \cdot p_{ac_2}$ as a function of $x$ for both LiPF$_6$ (top right panel) and LiBF$_4$ (bottom right panel) systems. We see that although the theory matches the simulation almost exactly for $p_{c_2a} \cdot p_{ac_2}$, it predicts the onset of percolation prematurely. We see that for both LiPF$_6$ and LiBF$_4$, the MD simulations predicts that an ion network will be formed between 0.2 and 0.3 ($0.2<x^*<0.3$, the grey shaded region in the right panel of Fig.~\ref{fig:emim_libf4_lipf6}). However, the theory predicts ion network formation when $p^*_{c_2a} \cdot p^{*}_{ac_2}=1/(f_{c_2}-1)/(f_a-1)$, as indicated by the intersection of the solid black curve with the dashed black line. This intersection occurs at roughly 0.13 for LiPF$_6$ and 0.15 for LiBF$_4$. This discrepancy is only moderate, and likely indicates that the ion clusters are not perfect obeying the mean-field Cayley trees assumption, i.e. they contain loops. When associations form loops, they do not add any ions to the cluster, and thus will not contribute to percolation. Thus, percolation is suppressed as more loops are formed. This idea is demonstrated quite clearly when comparing the bond percolation threshold on a diamond lattice in 3D (0.39~\cite{vyssotsky1961critical}) to the bond percolation threshold for a Bethe lattice with coordination number of 4 (0.25). Both lattices would have lattice sites with four neighbors, but the diamond lattice is periodic with a finite unit cell, and thus allows for cluster loops, and therefore, its percolation threshold is considerably higher. Confinement to lower-dimensional structures, such as planar arrangements near a surface, would further contribute to the formation of loops and suppression of Cayley tree clusters. 

\begin{figure}
\centering
\includegraphics[width=\textwidth]{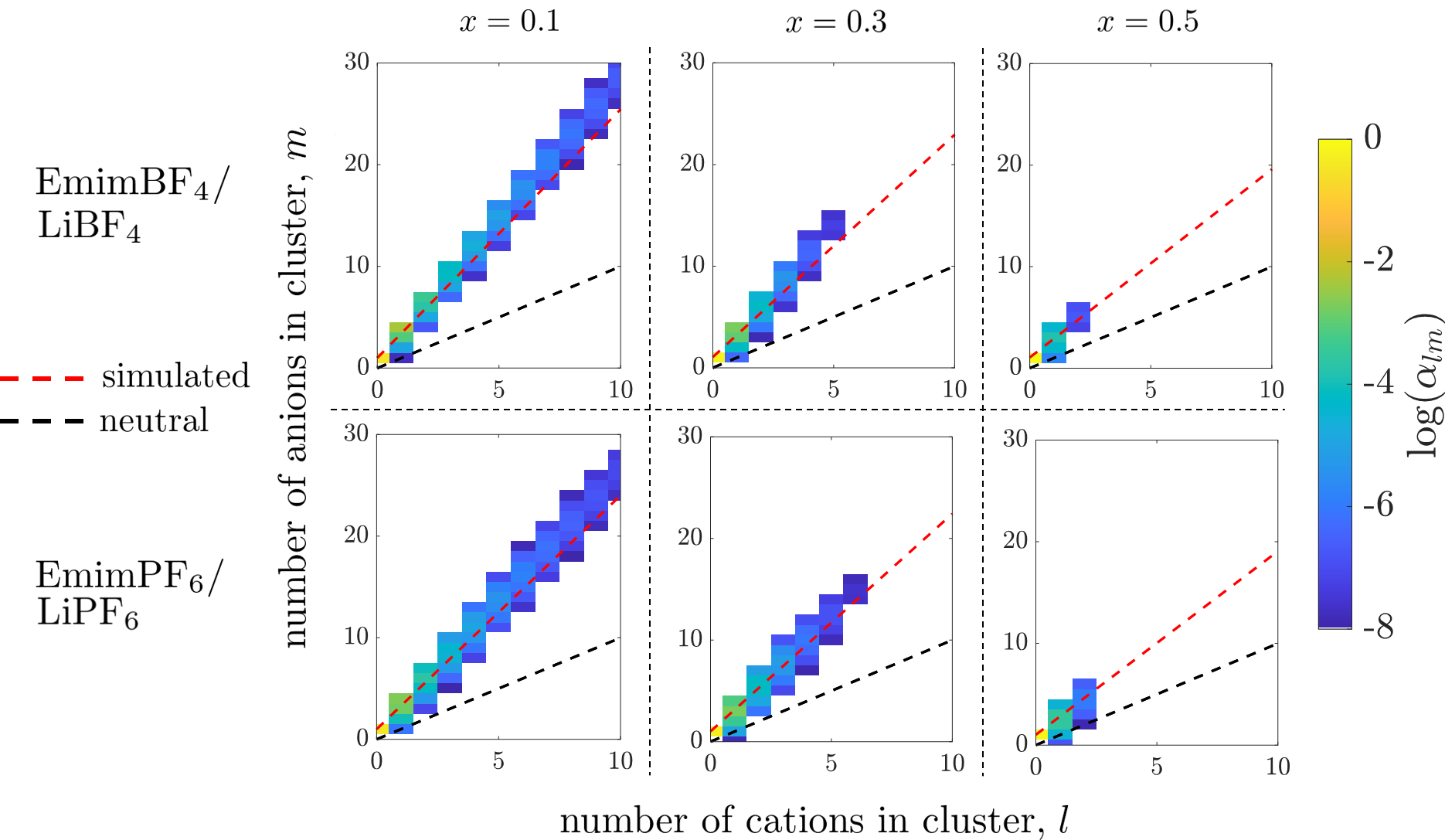}
\caption{The cluster distributions $\alpha_{lm}$, are plotted for emim/LiBF$_4$ (top panel) and emim/LiPF$_6$ (bottom panel) panel for lithium mole fractions of 0.1 (left panel), 0.3 (middle panel), and 0.5 (right panel). In each plot we draw curves corresponding to neutral clusters  (black dashed line), and linear regressions of the simulation date from Ref.~\citenum{molinari2019general} (red dashed line).}
\label{fig:clusters}
\end{figure}

As was speculated in Ref.~\citenum{gouverneur2018negative}, the formation of asymmetric ion clusters is at the core of the anomalous transport properties of alkali cations dissolved in ILs. This was confirmed via molecular simulations in Ref.~\citenum{molinari2019general}, where the distribution of clusters was shown to have an immense preference to form negative clusters. Such an observation is intuitive as the number of anions will outnumber the number of alkali cations in the mixture. In Fig.~\ref{fig:clusters}, we plot a color map of the theoretical \emph{sol} cluster distribution, $\alpha_{lm}$, for LiPF$_6$ and LiBF$_4$ salt in IL systems at various lithium salt mole fractions (left: 0.1, middle: 0.3, left: 0.5). Note that the \emph{sol} cluster distribution does not consider ions that are part of the ion network, and is defined simply as
\begin{align}
    \alpha_{lm}=\frac{(l+m)N_{lm}}{\sum_{lm}(l+m)N_{lm}}.
\end{align}
Within each plot we draw two lines: the black line corresponds to clusters with an equal number of lithium and anion ($l=m$) and the red line is a linear regression of the MD simulated clusters in Ref.~\citenum{molinari2019general}. We can see that in all cases, our theoretically predicted cluster distributions (using the parameters fitted to MD) are almost exactly in line with the linear regression of the simulated clusters. Thus, it is clear that our model is able to reproduce the cluster asymmetry found in MD simulations and speculated from experiments. Moreover, the theoretical cluster distributions are in agreement with the simulated distributions, not only in terms of asymmetry, but also in terms of decreasing in breadth beyond the percolation threshold. In other words, as the network increases in size smaller clusters tend to dominate the sol cluster distribution~\cite{mceldrew2020theory}. Such a trend is crucial in driving the effective charge reversal of alkali cations in these systems at larger salt fractions. 

\begin{figure}
\centering
\includegraphics[width=\textwidth]{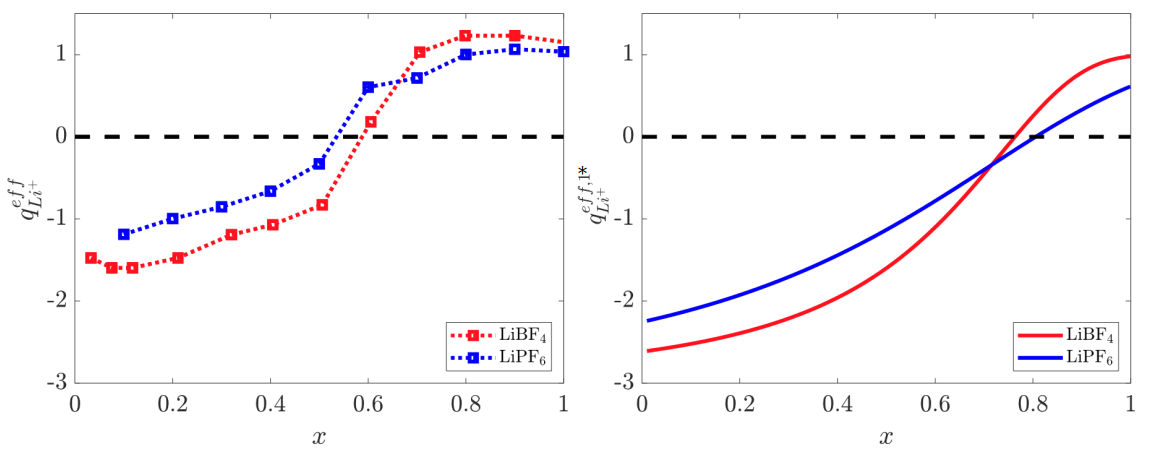}
\caption{The effective charge as computed from MD simulations performed in Ref.~\citenum{molinari2019general} (left panel) as well as Eq.~\eqref{eq:q1s} (right panel) with the theoretical parameters fitted from MD are plotted as function of lithium salt fraction for both emim/LiBF$_4$ (red) and emim/LiPF$_6$ (blue) systems.}
\label{fig:emim_libf4_lipf6_qeff}
\end{figure}

Finally, we compare the effective charge of lithium ions computed from the molecular simulations with the values predicted from Eq.~\eqref{eq:gen_eff_one_s} (using the previously derived probability equations with values for the parameters fitted to the simulations above). Overall, the theory prediction is in good qualitative agreement with the molecular simulations of Ref.~\citenum{molinari2019general}. At low mole fractions, we also predict there to be negative effective charges of the lithium ions. As the mole fraction of the lithium salt increases, from $x=0.1$ to $x=0.4$, the effective charge of the lithium cations slowly increases to less negative values. This occurs because as the mole fraction of the alkali metal salt increases, the large clusters aggregate into the gel phase, leaving the number of clusters in the sol with only a handful lithium cations relatively constant, as seen in Fig.~\ref{fig:clusters}. Upon reaching a mole fraction of $x=0.5$, further increases in $x$ cause a rapid increase in the effective charge of lithium. This occurs because at $x=0.5$, as seen in Fig.~\ref{fig:clusters}, there are practically only clusters in the sol with 1-2 lithium cations, and therefore, further increasing the mole fraction rapidly causes these highly negative clusters (which reside close to the red line in Fig.~\ref{fig:clusters}) to aggregate into the gel. Thus, leaving only small clusters and ion pairs in the sol. Further increasing the mole fraction beyond $x=0.7$ causes the effective charge of the lithium cations to recover its nominal charge of 1, owing to practically only free lithium cations existing in the sol phase at these mole fractions. 

The developed theory can even capturing the subtle differences in trends between the two systems (EmimBF$_4$/LiBF$_4$ and EmimPF$_6$/LiPF$_6$).
In simulations, lithium ions in the LiBF$_4$ system were expected to be effectively more negative at low lithium salt fractions than in the LiPF$_6$ system, but more effectively positive at high lithium salt fractions than in the LiPF$_6$ system. This phenomenology was exactly captured by the theory, and it can be traced back primarily to the difference in the association constants of the two salts. The LiBF$_4$ system was found to have $\ln \lambda_0=6.2$, while The LiPF$_6$ system was found to have $\ln \lambda_0=4.5$. Thus, BF$_4^-$ interacts with considerably more strength with Li$^+$ than PF$_6^-$. Thus, at low lithium salt fractions, prior to the formation of the ion network, lithium in the LiBF$_4$ system has a higher probability of being fully coordinated by anions than in the LiPF$_6$ system. Thus, prior to the formation of the ion network, increasing the number of associations decreases the effective charge of lithium. However, after the formation of the ion network, the associations become increasingly directed towards network. Thus, as more associations are formed, a higher fraction of those associations take place within the network, leaving ions in the sol more free~\cite{mceldrew2020theory}. The network is assumed not to contribute to the conduction of ionic current, and thus, it is the states of ions in the sol that matters in this case~\cite{mceldrew2020corr}. The difference in anion functionality also plays a secondary role in this regard as well. As $f_-$ increases relative to $f_+$, the system tends to partition more anions to the sol. When more anions are present in the sol, more associations take place in the sol, ultimately leaving lithium ions comparatively more bound. 

\begin{figure}
\centering
\includegraphics[width=0.8\textwidth]{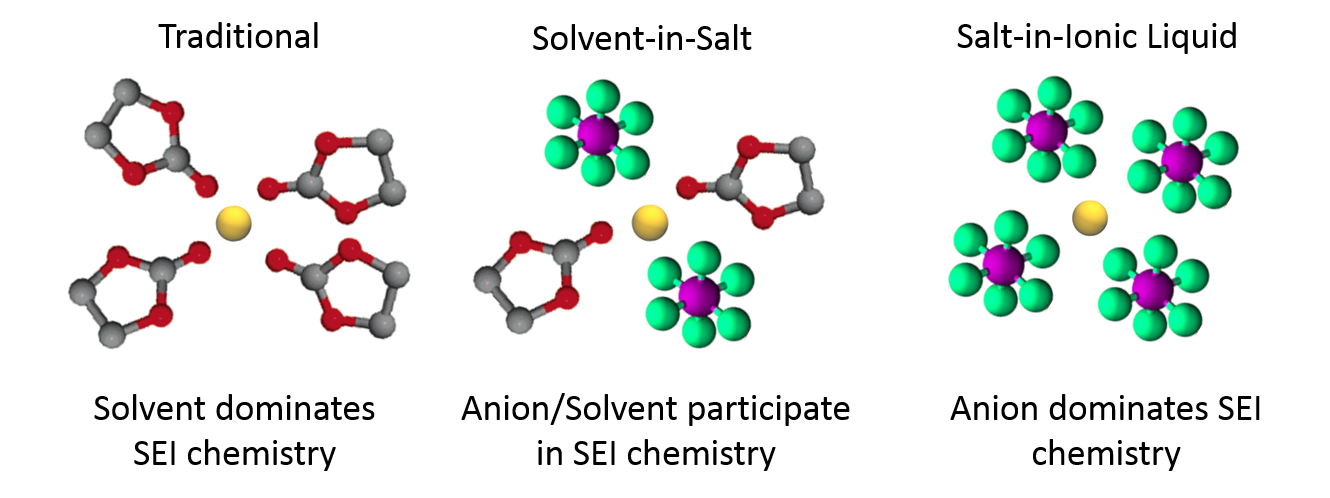}
\caption{A schematic of the expected coordination shell of lithium in traditional (left panel), solvent-in-salt (middle panel), and salt-in-IL (right panel) electrolytes. The example solvent shown is ethylene carbonate. }
\label{fig:coordination}
\end{figure}

While the qualitative trend of the effective charge of the lithium ions is in good agreement, we do not have a perfect quantitative match. We suspect the primary reason for the mismatch is that our definition for the effective charge is purely based on equilibrium ion association, whereas the definition for the effective charge for lithium from the simulations has some dynamical factors that contribute to its determination. One of the most important factors is the lifetime of the ion associations~\cite{mceldrew2020corr}. It was found in Ref.~\citenum{molinari2020chelation} that after roughly 1~ns, only about 40\% of the anions initially coordinated to lithium ions remain coordinated to lithium. Thus, the picture that lithium diffuses with its coordination shell intact is not strictly true. Moreover, the dissociation of anions coordinating lithium during diffusion would tend to increase the effective charge of lithium closer to its nominal charge of 1, which would likely push the theoretical predictions more in line with the effective charges computed from MD simulations. 

Interestingly, our theory suggests that there should be dynamic heterogeneity in the lithium transport, which should presumably change drastically with $x$. At small $x$, we expect there to be a broad distribution of dynamical states of lithium, owing to the cluster distribution being distributed over a wide number of possible clusters, as seen in Fig.~\ref{fig:clusters}. For $x$ closer to 0.5 and above, we expect the dynamical states of lithium to be more polarised. As practically only small clusters of lithium and the gel phase exist, it should be apparent that some lithium is free to move, but a significant proportion is trapped in the ionic network. In fact, dynamic heterogeneity of ion has been observed in ILs systems, notably by Hu and Margulis~\cite{Hu2006} and Feng \textit{et al.}~\cite{feng2019free}. Moreover, Araque \textit{et al.}~\cite{Araque2015} showed that this dynamic heterogeneity was a result of the strong correlations between ions because of their charge. Therefore, we expect dynamic heterogeneity of the lithium cations to be apparent.

The general trend of negative effective charges in salt-in-ionic liquid is problematic when considering their application in energy storage devices, especially in lithium-ion or lithium metal batteries. For one, in order to transport lithium through the electrolyte, the system will have to develop large concentration gradients to force lithium ions to generate a positive current. These concentration gradients translate into internal resistances that reduce the overall efficiency of the battery. Furthermore, as was demonstrated in Ref.~\citenum{bai2016transition}, limitations in the transport lithium ions drive the formation of dendritic lithium in lithium-ion batteries. Despite these shortcomings, IL based electrolytes have still gained a good deal of attention as candidate electrolytes in next-generation lithium-ion/lithium metal batteries (as well as sodium analogues). A primary reason for this is the enhanced battery safety obtained by replacing flammable organics with non-flammable ILs. However, another important attribute of salt-in-ionic liquid electrolytes is that they have actually shown high rate capabilities~\cite{yoon2013fast} in addition to dendrite suppression~\cite{schweikert2013suppressed,basile2016stabilizing} in lithium metal batteries. This occurs despite the presence of the aforementioned strong transport limitations that would seemingly \emph{reduce} rate capability and \emph{drive} the formation of dendrites. It seems an answer to this contradiction must lie within the ability of salt-in-IL electrolytes to form stable, low-resistance solid-electrolyte-interphases.

One concept reviewed thoroughly in Ref.~\citenum{li2020new} is that ion coordination governs the chemistry of the solid-electrolyte interphase (SEI), which is a passivating layer formed on the anode surface that enables ion intercalation/deposition reactions and prevents runaway decomposition of the electrolyte. This idea has been discussed extensively in the context of solvent-in-salt electrolytes \cite{yamada2019advances,li2020new}. In particular, as depicted in Fig.~\ref{fig:coordination}, in solvent-in-salt electrolytes lithium will be partially coordinated by anions in addition solvent molecules (shown as ethylene carbonate in Fig.~\ref{fig:coordination}), whereas in traditional battery electrolytes ($\sim$1~M) lithium ions will almost always be completely coordinated by solvent. This implies that in solvent-in-salt electrolytes, anions (in addition to the solvent) can participate in SEI chemistry. Thus, the chemistry driving the formation of SEI, can be very different at high salt concentrations in comparison to low salt concentrations where the chemistry of the SEI is dominated by the solvent. Furthermore, anion-derived SEI chemistries such as LiF have been leveraged to enable graphite intercalation, lithium plating, and high charge/discharge rates in systems were incapable of doing so at lower salt concentrations~\cite{yamada2019advances}. If we extrapolate the connection between lithium coordination and SEI chemistry to salt-in-IL systems, we can speculate that the anion cannot only participate in SEI chemistry, but it will likely \emph{dominate} SEI chemistry. In this case, it is likely that the anion-derived qualities of the SEI would be even more apparent in salt-in-IL systems. 

Thus, strong anion coordination is a ``double-edge sword" and salt-in-IL systems must balance the attributes gained from the anion-derived SEI with the losses incurred by sub-optimal lithium transport. Manipulating the ratio of IL salt to lithium salt, as well as incorporating solvent molecules~\cite{chen202063}, chelating agents~\cite{Ackermann2021,molinari2020chelation}, or co-anions ~\cite{Suo2016,wang2020highly} could serve as a design pathway for future concentrated electrolyte blends that interpolate in between the solvent-in-salt and IL regimes. The trouble is that as more components comprise the electrolyte formulation, the parameter space undergoes a combinatoric explosion. Therefore, theoretical models, such as ours, may become increasingly important in providing the molecular-level intuition required to navigate such a high-dimensional design space that experimentation and molecular simulation alone cannot cover. Moreover, our model could provide a convenient mathematical framework for accelerating physics-based machine learning of promising new electrolyte formulations, in which the model encodes some of the key features to be learned from the data, as has been used successfully in other areas of battery design~\cite{Muratahan2021,Adarsh2020}.

\section{Conclusion}

The statistical theory presented here, based on our well validated formalism for pure ionic liquids~\cite{mceldrew2020corr}, provides us with a clear picture of clustering and ionic network formation in alkali metal salt-in-ionic liquid systems. The take-home messages of this analysis are as follows:

\begin{enumerate}
    \item When alkali cations are dissolved in ionic liquids, they tend to be extensively coordinated by anions, and this has strong implications for some of their physico- and electro-chemical properties. Our model introduces a few key molecular parameters, principally the fixed ion functionalities, to capture changing ion coordination environments, which help understand the factors that control ion coordination in salt-in-IL electrolytes.
    \item At low alkali metal salt fractions, the majority of alkali cations exist in small, \emph{finite} negatively charged clusters. Our theory is able to quantitatively predict that negative effective charges of alkali cations are generally expected when alkali cations interact strongly with the anions, and the concentration of alkali cations is pre-critical.
    
    \item At high alkali metal salt fractions ($x$) beyond the critical gelation point, the electrolytes forms a percolating ion network. After this point, the ion associations are heavily directed towards the network, and the remaning ions in the sol become more free. Owing to the lack of mobility of the network, the sol ions dominate the conduction of current, despite their diminishing number as $x$ increases. Eventually, for high enough $x$, the probability of associations for ions in the sol becomes low enough that alkali cations in the sol diffuse without strong anion coordination, restoring their effective charge to the dilute value ($+1$) in the high concentration limit.
    
    \item Our model shows that the IL cation is not just an inert bystander in salt-in-IL electrolytes. It can interact favorably with the anion (which our theory models via the fitted $\chi$ parameter), and shift the equilibrium away from forming alkali-cation/anion clusters. This effect is most evident at low $x$ when the IL cation is abundant. In effect, the IL cations ``compete" with alkali cations for interactions with the anions. The result is that as $x$ increases (and IL cation concentration decreases) alkali-cation/anion clustering is strongly promoted.
    
\end{enumerate}



While not explored in great detail in this work, our model can be leveraged to help find the optimal concentration of a salt-in-ionic liquid electrolyte which must balance several competing factors. First, the mobility of alkali cations is a product of their quantity and effective charges, as well as the inverse viscosity of the mixture. This implies that mobility of alkali cations will not be maximized until after its effective charge becomes larger than 0. Thus, for the Emim/LiBF$_4$ and Emim/LiPF$_6$ systems, the mobility of alkali cations will be maximized at very high alkali metal salt fractions ($x>0.5)$ after the system has formed a percolating ion network and the effective charge of lithium has recovered positive values. This is significantly higher than what might be intuitively expected for these systems, because the viscosity of the mixture drastically increases with $x$. Second, we must remember that although the transport of alkali cations is hindered by anion coordination, the formation of stable, low resistance SEI layers can be greatly enhanced by anion coordination. Thus, we do not want to blindly optimize these electrolytes for alkali-cation mobility without considering the repercussions for anion-derived SEI chemistries. Therefore, if our design strategy is to increase alkali-cation mobility by reducing its association to anions (perhaps via the introduction of solvents, chelating agents, or less associative anions), then we should be careful to not dilute our electrolyte so much that the coordination shells of alkali cations begin to look more like those in traditional solvent-dominated systems. 

Currently, negative transference numbers have been detected experimentally in multiple systems via electrophoretic NMR measurements of ion mobility. Moreover, MD simulations have shown the formation of percolating ion networks in salt-in-IL electrolytes, as well as other super-concentrated systems. Furthermore, researchers have observed that salt-in-IL electrolytes seem to form a ``translucent gel" above certain alkali metal salt fractions, which could be evidence that a percolating ion network has formed in the system~\cite{reber2021anion}. However, the coupling of ion mobility to percolating network formation has not been unequivocally demonstrated in experiments. Motivated by our statistical theory, future experiments could perhaps shed light on this phenomena, by combining measurements of ion mobility and conductivity, with a systematic rheological detection of gelation~\cite{winter1986analysis}. Alternatively, high-energy X-ray total scattering~\cite{fujii2017long} or small-angle x-ray scattering~\cite{qian2021insights} could potentially provide experimental verification of ion network formation via comparisons of simulated and measure structure factors.

\section{Acknowledgements}

MM and MZB acknowledge support from a Amar G. Bose Research Grant. This work used the Extreme Science and Engineering Discovery Environment (XSEDE), which is supported by National Science Foundation grant number ACI-1548562. ZG was supported through a studentship in the Centre for Doctoral Training on Theory and Simulation of Materials at Imperial College London funded by the EPSRC (EP/L015579/1) and from the Thomas Young Centre under grant number TYC-101. AK would like to acknowledge the research grant by the Leverhulme Trust (RPG-2016- 223). We would like to acknowledge the Imperial College-MIT seed fund. NM is supported by the US Department of Defense MURI under Award No. N00014-20-1-2418 and used the computing resources provided by the Harvard University FAS Division of Science Research Computing Group.

\bibliography{wise.bib}

\end{document}